\documentclass[12pt,letterpaper]{article}

\usepackage[ left=1in, top=1in, right=1in, bottom=1in]{geometry}
\usepackage{amsmath,amsfonts,amssymb,bbm}
\usepackage{amsthm,colonequals,url}
\usepackage{cite}
\usepackage[sort]{natbib}
\usepackage{xr} 
\usepackage{graphicx}
\usepackage{xcolor}
\newcommand{\mycellcolor}{\leavevmode\color{gray}}

\newtheorem{definition}{Definition}
\newtheorem{theorem}{Theorem}
\newtheorem{corollary}{Corollary}
\newtheorem{lemma}{Lemma}

\newcommand{\one}[1]{\ensuremath{\mathbbm{1}}(#1)}
\newcommand{\dd}{\, {\rm d}}
\newcommand{\zz}{\mathbb{Z}}
\newcommand{\real}{\mathbb{R}}

\newcommand{\myE}{\mathbb{E}}
\newcommand{\myP}{\mathbb{P}}

\newcommand{\cA}{\mathcal{A}}
\newcommand{\cF}{\mathcal{F}}
\newcommand{\cG}{\mathcal{G}}
\newcommand{\cI}{\mathcal{I}}
\newcommand{\cL}{\mathcal{L}}
\newcommand{\cN}{\mathcal{N}}
\newcommand{\cP}{\mathcal{P}}

\newcommand{\hth}{\hat{\theta}}
\newcommand{\thh}{\hat{\theta}_T}
\newcommand{\thetatrue}{\theta_*}

\providecommand{\keywords}[1]{\textbf{\textit{Keywords---}} #1}

\title{Interpretation of point forecasts with unknown directive}

\author{Patrick Schmidt\footnote{\emph{Address for
			correspondance:} Patrick Schmidt, HITS gGmbH,
		Schloss-Wolfsbrunnenweg 35, 69118 Heidelberg, Germany. E-mail:
		Patrick.Schmidt@h-its.org.}\\
	Heidelberg Institute for
	Theoretical Studies (HITS), Heidelberg, Germany and \\ Goethe University
	Frankfurt, Frankfurt, Germany \and
	Matthias Katzfuss \\Texas A\&M University, College
	Station, USA \and
	Tilmann Gneiting \\Heidelberg Institute for
	Theoretical Studies, Heidelberg, Germany and \\ Karlsruhe Institute
	of Technology (KIT), Karlsruhe, Germany\\\author{Patrick Schmidt \and Matthias Katzfuss \and Tilmann Gneiting}}

\begin{document}
	\maketitle

\begin{abstract}
	
	Point forecasts can be interpreted as functionals (i.e., point
	summaries) of predictive distributions.  We consider the situation
	where forecasters' directives are hidden and develop methodology for
	the identification of the unknown functional based on time series data
	of point forecasts and associated realizations.  Focusing on the
	natural cases of state-dependent quantiles and expectiles, we provide
	a generalized method of moments estimator for the functional, along
	with tests of optimality relative to information sets that are
	specified by instrumental variables.  Using simulation, we demonstrate
	that our optimality test is better calibrated and more powerful than
	existing solutions.  In empirical examples, Greenbook gross domestic
	product (GDP) forecasts of the US~Federal Reserve and model output for
	precipitation from the European Centre for Medium-Range Weather
	Forecasts (ECMWF) are indicative of overstatement in anticipation of
	extreme events.
	
\end{abstract}

\keywords{expectile, generalized method of moments, identifying moment
	conditions, information set, loss function, optimality of point forecasts,
	quantile}

\section{Introduction}  \label{sec:introduction} 

Forecasts are frequently the basis of crucial decisions.  Yet, they
are fraught with uncertainty due to imperfections in the observation,
understanding, and modeling of the underlying mechanisms.  To account
for this uncertainty, it is increasingly being recognized that
forecasts ought to be probabilistic in nature \citep{Gneiting2014}.
If forecasts are issued in the form of full predictive distributions,
there are well established methods for computing the Bayes act in any
given decision problem, for testing optimality, and for comparing and
ranking competing forecasting methods.

However, single-valued point forecasts remain ubiquitous.  Their
interpretation requires assumptions on the decision process or
directive that the forecasters used in order to generate the point
predictions \citep{Elliott2008a, Engelberg2009, Manski2016}.  A
directive can be expressed through a functional (i.e., a real-valued
summary) of the predictive distribution, and it is a widely used
assumption that the reported functional is the mean value or
expectation. However, there is often little justification for this
supposition. Knowledge of the functional used to generate
the forecast is important, as it allows for proper interpretation,
evaluation, testing, and comparison of point predictions
\citep{Gneiting2011}.

\begin{figure}	
	
	\centerline
	{
		\includegraphics[width=0.95\textwidth]{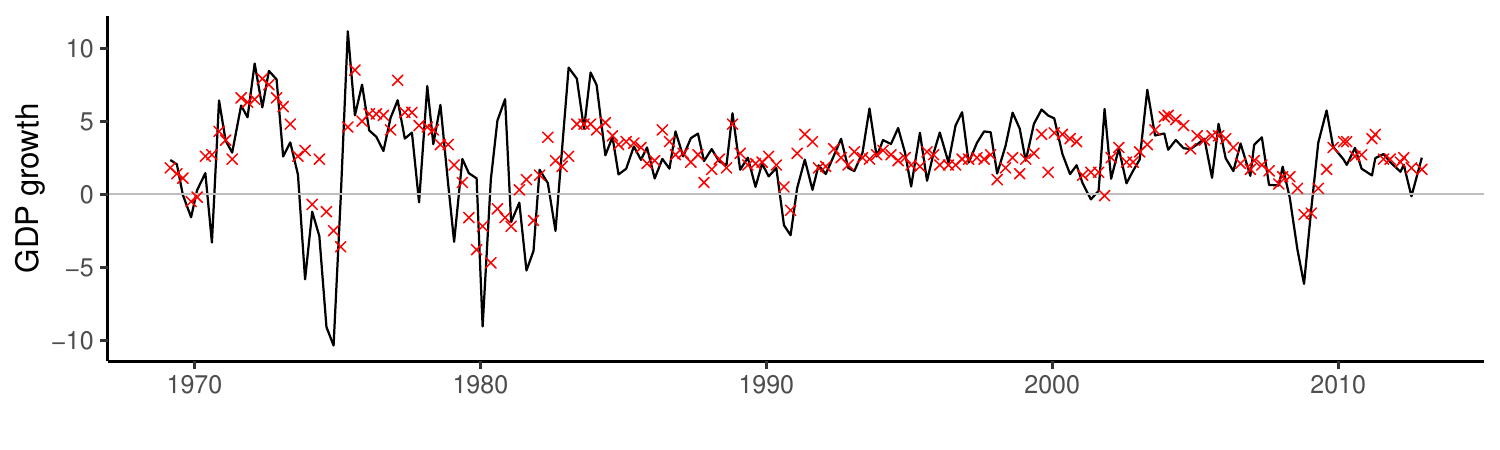}
	}	
	
	\vspace{-3mm}
	
	\caption{Time series of one-quarter ahead Greenbook forecasts
		(crosses) and respective observations (solid line) of real GDP
		growth in the US (in percent).  \label{fig:data_gdp}}
	
\end{figure}

\begin{figure}	
	
	\centerline
	{
		\includegraphics[width=0.95\textwidth]{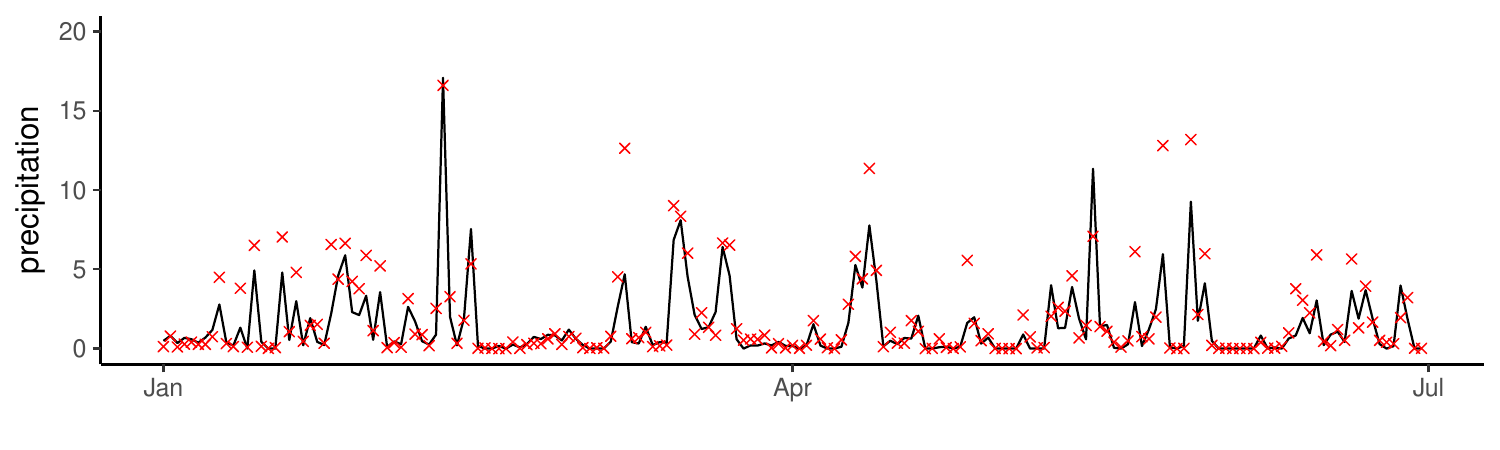}
	}	
	
	\vspace{-3mm}
	
	\caption{Time series of 24-hour ahead ECMWF forecasts (crosses) and
		respective observations (solid line) of daily accumulated
		precipitation (in millimeter) at London, UK in 2013.
		\label{fig:data}}
	
\end{figure}

Here we address the setting of point forecasts with unknown directive,
for which the forecaster implicitly (only) reports a certain
functional of the predictive distribution.  This scenario can arise
with expert forecasts or response items in surveys.  Another important
example is output from complex computer models, such as in weather and
climate prediction, which are often tuned by multiple individuals to
achieve forecasts that model developers or decision makers perceive as
optimal, in ways that might neither be transparent nor explicitly
defined.  Such forecasts would be most informative if the user knew
the directive under which the forecast was issued.  Our goal here is
to estimate the functional from time series data of point forecasts
and associated realizations, and to construct tests regarding the
properties of the functional.  The type of data encountered in this
setting is illustrated in Figure \ref{fig:data_gdp}, which displays
one-quarter ahead Greenbook gross domestic product (GDP) forecasts of
the US~Federal Reserve, and in Figure \ref{fig:data}, which shows
24-hour ahead forecasts of daily accumulated precipitation at London,
UK from the high-resolution run operated by the European Centre for
Medium-Range Weather Forecasts (ECMWF).  Once the functional has been
estimated, the point forecasts can be coherently interpreted, and be
compared to other point or probability forecasts, and constructive
feedback can be given to model developers.

Extant work on estimating a directive based on time series data of
point forecasts and realizations has focused on the estimation of loss
functions.  The pioneering work of \citet{Elliott2005} provides a
generalized method of moments (GMM) estimator of the loss function
under constant preferences and linear forecasting models.
\citet{Patton2007} apply methods of this type to the US~Federal
Reserve's Greenbook GDP forecasts with a new class of loss functions,
which consists of quadratic splines that depend on a state variable in
flexible ways.  Recently, asymmetric piecewise linear and piecewise
quadratic loss functions have been estimated in various economic
applications \citep{Christodoulakis2008, Capistran2008, Elliott2008b,
	Krol2013, Pierdzioch2013, Wang2014, Fritsche2015, Guler2017}.  In a
neuroscience application, \citet{Kording2004} estimate the loss
function implicit in human sensorimotor control by varying targets in
an experimental task.  \citet{Sims2015} uses a similar approach to
infer the implicit loss function of the visual working
memory. \citet{Lieli2013} discuss the recoverability of the loss
function from theoretical perspectives.

Here we argue that the loss function is not identifiable from point
forecasts and realizations only.  Hence, we formalize notions of
optimality for point forecasts in terms of functionals rather than
loss functions.  We consider single-valued scalar functionals
throughout, although the results extend to set-valued and multivariate
cases under additional technical considerations.  For estimation, we
focus on parametric models for quantiles and expectiles, where the
level of asymmetry depends on state variables.  We propose a GMM
estimator and apply standard GMM theory to show consistency and
asymptotic normality under mild assumptions.  We also discuss testing
of forecast optimality and other forecast properties.  This
generalizes the approach of \citet{Elliott2005} to state-dependent
settings.  In comparison to \citet{Patton2007} our methods yield
improved interpretability and nominally sized tests.

Importantly, forecast optimality needs to be defined and studied
relative to information sets \citep{Holzmann2014}.  The choice of
instrumental variables for the GMM estimator determines the
information set and, therefore, the mode of optimality that is tested
for.  Adding information into the instrument vector yields stronger
hypotheses of forecast optimality and increases power against
suboptimal forecasts. In our real data and simulation examples we
elucidate the role of the information set, and we illustrate that the
proposed optimality test can be used to distinguish information bases.

The remainder of the paper is organized as follows.  In Section
\ref{sec:identification}, we review notions of optimality for point
forecasts and discuss the identifying moment conditions upon which our
approach is based.  In Section \ref{sec:estimation}, we introduce a
parametric GMM estimator in the time series setting, study its large
sample behavior, and discuss tests of optimality and more specific
hypotheses.  The Monte Carlo studies in Section \ref{sec:simulation}
serve to compare our approach to existing solutions.  Section
\ref{sec:empirical} is dedicated to empirical studies, where our
approach yields accessible and scientifically relevant insights.  In
particular, we posit that Greenbook GDP forecasts of the US~Federal
Reserve and ECMWF forecasts of daily accumulated precipitation can be
interpreted as state-dependent quantiles or expectiles, and we observe
a pronounced tendency to exaggerate in anticipation of tail events.
The paper closes with a discussion in Section \ref{sec:discussion}.
Technical results and proofs are provided in Appendices
\ref{app:ident} to \ref{app:asnormal}.  An online supplementary
material document contains additional detail in Sections
\ref{supp:exasymm} through \ref{supp:statedependent}.

\section{Optimality of point forecasts and identifying moment conditions}  \label{sec:identification}

Consider a real-valued random variable $Y$ and a corresponding point
forecast $X$, which is based on the information available to the
forecaster, as encoded by some $\sigma$-algebra $\cF$.  Commonly, a
point forecast is interpreted as the mean of the conditional
distribution $\cL(Y|\cF)$, i.e.,
\[
X = \myE[Y|\cF].
\]
Here and throughout the paper, equality of random variables is
understood to hold almost surely.  Proceeding to a more general
framework, let $\alpha : \cP \mapsto \real$ be a functional, i.e., a
single-valued mapping from some class $\cP$ of probability
distributions to the real line (\citeauthor{Horowitz2006},
\citeyear{Horowitz2006}; \citeauthor{Huber2009}, \citeyear{Huber2009},
p.~9).  A functional $\alpha$ is symmetric if, for every symmetric
distribution $P \in \cP$ with symmetry point $c$, it holds that $c =
\alpha(P)$.  Prominent alternatives to the mean functional are symmetric
functionals, like the median, or asymmetric generalizations, such as
quantiles, expectiles, and generalized quantiles \citep{Bellini2014}.
Throughout, we use the short notation $\alpha(Y|\cF)$ for
$\alpha(\cL(Y|\cF))$.

\begin{definition}[optimal $\alpha$-forecast]  \label{def:optimal}
	A random variable\/ $X$ is an optimal $\alpha$-fore\-cast of\/ $Y$
	relative to the information set\/ $\cF$ if
	\[
	X = \alpha(Y|\cF).
	\] 
\end{definition}

Now, crucially, we consider the situation in which the functional used
by the forecaster \emph{and}\/ the conditional distributions
$\cL(Y|\cF$) are unknown.  In line with seminal work on professional
economic forecasters \citep{Elliott2005, Patton2007}, we merely assume
that the unknown conditional distribution constitutes a predictive
distribution consistent with some information set $\cF$.  

Next we introduce notation.  If $R$ is a random variable or random
vector and $Q$ is an integrable random variable, the relation $R \in
\cF_Q$ indicates that $R$ is $\cF$-measurable and both $R$ and $QR$
are (componentwise) integrable.  The relation $R \in \cF$ means that
$R$ is $\cF$-measurable and integrable.  As usual, we write
$\sigma(W)$ for the information set generated by the random vector
$W$.  Finally, the partial derivative of a function $g(x,y)$ with
respect to $x$ is denoted $g_{(x)}(x,y)$.

Standard properties of conditional expectations \citep[e.g.,][Section
34]{Billingsley} yield identifying moment conditions for an optimal
mean-forecast relative to the information set $\cF$, namely,
\[
X = \myE[Y|\cF] \iff \myE[(X-Y) W] = 0 \; \text{ for all } \; W \in
\cF_{X-Y},
\]
where the components of the random vector $W$ (henceforth called
\emph{instruments}) represent information $\cF$ available to the
forecaster when the prediction is issued.  This property of optimal
mean-forecasts generalizes to optimal $\alpha$-forecasts.
Specifically, for every sufficiently regular functional $\alpha : \cP
\mapsto \real$ there exists a function $V_\alpha$ identifying the
optimal $\alpha$-forecast, i.e.,
\begin{equation}  \label{eq:ident}
	X = \alpha(Y|\cF) \iff 
	\myE [V_\alpha(X,Y) W] = 0 \; \text{ for all } \; W \in \cF_{V_\alpha(X,Y)}.
\end{equation}
We refer to \eqref{eq:ident} as the \emph{identifying moment
	conditions}\/ for an optimal $\alpha$-forecast and give rigorous
versions thereof in Appendix \ref{app:ident}.  The identification
function $V_\alpha$ is unique up to an $\mathcal{F}$-measurable
multiplicative factor \citep[Thm.~8]{Steinwart2014}, so the set of the
arising moment conditions does not depend on its choice.  The moment
conditions are the foundation of the methodology developed
hereinafter.  In particular, they allow for tests of whether a point
forecast is an optimal $\alpha$-forecast relative to a specific
information set.  In the hypothetical limit of an infinite supply of
data and instruments, a non-rejection of the test is a sufficient
condition for optimality.

In contrast to our approach, extant work \citep[e.g.,][]{Elliott2005}
has typically defined optimal point forecasts via a loss function $L$,
namely as
\[
X = \arg {\textstyle \min_{x \in \real}} \, \myE[L(x,Y) | \cF].		
\]
Under regularity conditions, this specifies a well-defined optimal
$\alpha_L$-forecast, where the functional $\alpha_L$ is defined as
\[
\alpha_L : \cP \mapsto \real, \quad P \mapsto \arg {\textstyle \min_{x \in \real}} \,
\myE_{Y \sim P}[L(x,Y)]
\]
for a suitable class $\cP$ of probability distributions.  For example,
the mean-functional can be defined as the minimizer of expected
quadratic loss, $L(x,y) = (x-y)^2$, for probability distributions with
finite second moments.  While some functionals, such as the expected
shortfall and the mode, do not admit definitions via loss functions
relative to broad classes of probability distributions
\citep{Gneiting2011, Heinrich2014}, every functional $\alpha$ with
identifying moment conditions \eqref{eq:ident} can be defined via a
loss function $L$ under weak conditions, and the identification
function $V_\alpha$ derives from the partial derivative $L_{(x)}(x,y)$
(\citeauthor{Steinwart2014}, \citeyear{Steinwart2014}, Thm.~8;
\citeauthor{Fissler2016}, \citeyear{Fissler2016}, Thm.~3.2).  In this
setting, the literature commonly states necessary (but not sufficient)
conditions for optimality \citep[e.g.][]{Diebold1996, Patton2007,
	Patton2010}.  As a notable exception, Proposition 1 of
\citet{Elliott2005} supplies identifying moment conditions of the
general form \eqref{eq:ident} under piecewise linear and piecewise
quadratic loss.

Our rationale for the shift of emphasis from loss functions to
functionals is an identifiability problem.  Specifically, if a
functional is defined via a loss function, then there exists a whole
class of non-trivially distinct loss functions that define the very
same functional \citep{Steinwart2014, Ehm2016}.  It is therefore
futile to identify the shape of the loss, given that all these
functions lead to the same functional-forecast and moment conditions.
For example, while the mean-functional minimizes expected quadratic
loss, given any convex and differentiable function $\varphi$ the loss
function $L(x,y) = \varphi(y) - \varphi(x) - \varphi'(x)(y-x)$ also
induces an optimal mean-forecast \citep{Savage1971}.  In this light, a
more compelling approach is to estimate functionals.

To summarize, while loss functions are not identified, functionals are
identified to the extent that they differ on the class of the arising
conditional distributions $\cL(Y|\cF)$.  For instance, the mean and
the median are distinct functionals in general, but if all considered
distributions are symmetric, the two functionals are identical and
cannot be identified.  In the sequel we focus on optimal forecasts in
the form of either quantiles or expectiles, to allow for unique
identification without imposing unduly strong assumptions on the
data-generating process.

\section{Parametric estimation and testing of state-dependent quantiles and expectiles} 
\label{sec:estimation}

We turn to parametric estimation of possibly time-varying functionals.
Consider a stochastic process $(X_t,Y_t,Z_t)_{t \, \in \, \zz}$
of forecasts, observations, and covariates, for which we have a sample
path $(x_t, y_t, z_t)_{t \, = \, 1, \ldots, T}$.  Our goal is to infer
the functional represented by the point forecasts.

We assume that at each point in time an optimal forecast is issued,
i.e.,
\[
X_t = \alpha_t(Y_t|\cF_t) 
\]
for $t \in \zz$.  In the situation of an $h$-step ahead forecast, the
available information is typically generated by lagged variables of
the outcome and the vector-valued covariate, so that $\cF_t = \sigma(
\{ Y_u, Z_u : u \leq t - h \} )$.  For ease of notation, statements
about all time points are often denoted without subscripts.  For
example, we write $X = \alpha(Y|\cF)$ instead of $X_t =
\alpha_t(Y_t|\cF_t)$ for $t \in \zz$.

Extending Definition \ref{def:optimal}, we allow the functional
$\alpha$ to depend on the current situation, represented by some
$\cF_t$-measurable state variable $S_t$, for a
\emph{state-dependent}\/ functional.  For example, in the
aforementioned situation of an $h$-step ahead forecast $S_t$ might
include the most recent observation, $Y_{t-h}$, components of the
covariate vector $Z_{t-h}$, or the current forecast, $X_t$.
Asymmetric and state-dependent point forecasts can arise for a variety
of reasons, including varying preferences of the forecaster,
asymmetric information, and non-linear transformations of the data,
and we refer to Supplementary Section \ref{supp:exasymm} for details.
In the following we assume that the true functional is a
state-dependent quantile or state-dependent expectile of the
conditional distribution $\cL(Y|\cF)$.  As elaborated in Supplementary
Section \ref{supp:existence}, this assumption is of surprising
flexibility and generality.

\subsection{State-dependent quantiles and expectiles}  \label{sec:qe}

The $\tau$-quantile functional $q_\tau(P)$ of a distribution $P$ with
continuous and strictly increasing cumulative distribution function is
the unique solution $x$ to the equation $P((-\infty, x]) =
\tau \in (0,1)$. In our setting we can express this directly in
terms of the identification function of the $\tau$-quantile, namely
$V_\tau(x,y) = \one{y \le x} - \tau$:
\[
X = q_\tau(Y|\cF) \iff \myE[(\one{Y \le X} - \tau) \, W] = 0 \; \text{
	for all } \; W \in \cF.
\]
For technical details see Appendix \ref{app:ident}. 

While quantiles are asymmetric generalizations of the median,
expectiles are asymmetric generalizations of the mean.  Specifically,
\citet{Newey1987b} introduced the $\tau$-expectile $e_\tau(P)$ of a
non-degenerate distribution $P$ with finite mean as the
unique solution $x$ to the equation
\[
\frac{\tau}{1-\tau} = \frac{\int_{-\infty}^x(x-y) \dd
	P(y)}{\int_x^\infty(y-x) \dd P(y)},
\]
where $\tau \in (0,1)$. In our setting this is equivalent to
\[
X = e_\tau(Y|\cF) \iff \myE[|\one{Y \le X} - \tau| \, (Y-X) \, W] = 0
\; \text{ for all } \; W \in \cF_{X-Y},
\]
which reveals the corresponding identification function $V_\tau(x,y) =
|\one{y \le x} - \tau| (y-x)$.

We allow for additional flexibility and let the level $\tau$ of the
quantile or expectile functional depend on the state $s$ via a
parametric function $m(s,\theta)$.

\begin{definition}[specification model]	
	Let $\Theta$ be a subset of $\real^p$ and suppose that the state
	variable $s$ takes values in $\real^k$. A specification model is a
	measurable function $m(s,\theta)$ that maps $\real^k \times \Theta$
	into the unit interval $(0,1)$.
\end{definition}	

We say that a specification model $m(s,\theta)$ is continuous
(continuously differentiable) if it is continuous (continuously
differentiable) in $\theta \in \Theta$ for every $s \in \real^k$.
Examples for specification models are given in Table
\ref{tab:specification}, where the state $s$ is assumed to be
real-valued.  The \emph{constant}\/ model assumes that the forecaster
always states the $\theta$-quantile or expectile and has been
implemented in much previous work \citep{Elliott2005,
	Christodoulakis2008, Krol2013, Pierdzioch2013, Fritsche2015}.
Throughout we consider specification models that nest the constant
model and, therefore, include optimal quantile or expectile forecasts
as a special case.

\begin{table}
	
	\caption{Specification models for a real-valued state variable $s$.  \label{tab:specification}} 
	
	\begin{tabular}{lll}
		Name     & Model & Parameter Space $\Theta$ \\
		\hline
		Constant & $m(s, \theta) = \theta$ & $(0,1)$ \\
		Break     & $m(s, \theta) = \Phi(\one{s \leq \theta_0} \theta_1 +  \one{s > \theta_0} \theta_2)$ & $\real^3$ \\
		Linear     & $m(s, \theta) = \Phi(\theta_0 + \theta_1 s)$ & $\real^2$ \\
		Periodic & $m(s, \theta) = \Phi(\theta_0 + \theta_1 \sin(2\pi s/\theta_2))$ & $\real^2 \times (0,\infty)$ 
	\end{tabular}
	
\end{table}

For the state-dependent models we employ the probit link, where the
CDF $\Phi$ of the standard normal distribution ensures a quantile or
expectile level in the unit interval.  Alternatively, the logit link
or a related formulation could be used.  The \emph{break}\/ model
allows for structural change at a threshold, and the \emph{linear}\/
model admits linear dependence within the argument of the link
function.  The \emph{periodic}\/ model specifies the base level
$\theta_0$, the amplitude $\theta_1$, and the period $\theta_2$ of any
cyclic component.  Suitable adaptations apply when the state variable
$s = (s_1, \ldots, s_k)'$ takes values in $\real^k$.  For example, the
linear model generalizes to $m(s, \theta) = \Phi(\theta_0 + \theta_1
s_1 + \cdots + \theta_k s_k)$. In any practical problem, the choice of
the state variable and the parametric specification model will be
guided and informed by substantive expertise.

\subsection{Generalized method of moments (GMM) estimator}  \label{sec:GMM}

Let us now assume that $X_t$ is an optimal quantile-forecast with
state-dependent level $S_t$ prescribed by the specification model
$m(s, \thetatrue)$, i.e.,
\[
X_t = q_{m(S_t, \thetatrue)}(Y_t|\cF_t) 
\]
for $t \in \zz$.  Crucially, we assume that the state variable $S_t$
is $\cF_t$-measurable, so that
\[
\myE[(\one{Y_t \le X_t} - m(S_t, \thetatrue)) W_t] = 0 \; 
\text{ for all } \; W_t \in \cF_t. 
\]
In practice, we choose a $q$-variate instrument vector $w_t =
(w_{t,1}, \ldots, w_{t,q})'$ that comprises information available to
the forecast issuer at the time the forecast was made.  We refer to
\[
g_t(\theta) = (\one{y_t \le x_t} - m(s_t,\theta)) \, w_t
\]
as the \emph{moment function}, which we interpret as a mapping from
the parameter space $\Theta$ to $\real^k$.  Alternatively, $X_t =
e_{m(S_t, \thetatrue)}(Y_t|\cF_t)$ might be an optimal state-dependent
expectile-forecast, with associated moment function
\[
g_t(\theta) = |\one{y_t \le x_t} - m(s_t,\theta))| \, (x_t - y_t) \, w_t.
\]

Given a sample path of instrument vectors $w_1, \ldots, w_T$, the
empirical mean of the moment function is given by
\[
g_T(\theta) = \frac{1}{T} \sum_{t=1}^{T} g_t(\theta).
\]
The GMM estimator $\thh$ is obtained by minimizing the quadratic norm
of $g_T(\theta)$, in that
\begin{equation}  \label{eq:GMM}
	\thh = \arg {\textstyle \min_{\theta \in \Theta}} \, g_T(\theta)' \, M_T \, g_T(\theta)
\end{equation}
with a weighting matrix $M_T \in \real^{q \times q}$.  Subject to 
customary assumptions, which include continuity of the specification
model and sufficiently rich instruments to guarantee unique
identifiability, the GMM estimator $\thh$ in \eqref{eq:GMM} is
consistent.  For formal statements see Appendix \ref{app:consistency},
where we draw on standard GMM theory \citep{Hansen1982} and discuss
identification through appropriate choices of the instruments.

Once consistency has been established, asymptotic laws in GMM theory
can be applied. Throughout this study, we employ the standard two-step
GMM procedure, where in a first step $\thh$ is computed with the
identity matrix as weighting matrix and in a second step with the
inverse of the heteroskedasticity and auto\-correlation consistent
(HAC) covariance estimator $\Sigma_T$ based on the first step
estimate.\footnote{Specifically, we follow \cite{Newey1987a} and use
	the linear Bartlett kernel and a sample size dependent bandwidth of
	$m(T)= \lceil{T^{1/5}}\rceil$ such that
	\[
	\Sigma_T = \Omega_0 
	+ \sum_{j=1}^{m(T)} \left( 1 - \frac{j}{m(T)+1} \right) 
	\left( \Omega_j + \Omega_j' \right), \quad 
	\Omega_j = \frac{1}{T} \sum_{t=j+1}^{T} g_t g_{t-j}.
	\]
	Our empirical applications and simulation results are robust to
	data-dependent bandwidth selection as proposed in \cite{Andrews1991}
	and \cite{newey1994automatic}.}  Subject to regularity conditions that
include continuous differentiability of the specification model and
mixing conditions on the moment function, the two-step GMM estimator
is asymptotically normal with
\begin{equation}  \label{eq:asymnormal}
	\sqrt{T} \hspace{0.2mm} (\thh - \thetatrue) 
	\to \cN_p \! \left( 0, (G \, \Sigma^{-1} G')^{-1} \right) 
	\; \text{ as } \; T \to \infty,
\end{equation}
where $p$ is the dimension of the parameter vector, $\Sigma \in
\real^{q \times q}$ is the covariance matrix of the moment function,
and $G \in \real^{p \times q}$ is the expectation of its partial
derivative with respect to $\theta$, at the true parameter
value $\thetatrue$.  In the case of quantiles, we have $G = \myE[
m_{(\theta)}(S,\thetatrue) W']$ and
\[
\Sigma = \myE[(\one{Y \leq X} - m(S,\thetatrue))^2 \, W W'], 
\] 
and in the case of expectiles it holds that $G = \myE
[m_{(\theta)}(S,\thetatrue) \, |Y-X| \, W']$ and
\[
\Sigma = \myE[(\one{Y \leq X} - m(S,\thetatrue))^2 \, (Y-X)^2 \, WW']. 
\] 
For a formal treatment based on classical GMM theory
\citep{Hansen1982} see Appendix \ref{app:asnormal}.

While the GMM estimator itself is robust with respect to the selection
of the instruments if the specification model is identified, the
associated tests of forecast optimality depend heavily on the
instrument vector $W$, which determines the information set that is
tested for.  We turn to this issue now.

\subsection{Testing optimality with unknown directive}  \label{sec:testing}

The well-known test of overidentifying restrictions \citep{Hansen1982}
can be used to test forecast optimality.  Specifically, if the
dimension $q$ of the instrument vector $W$ is greater than the
dimension $p$ of the parameter vector $\theta$, and subject to the
same regularity conditions as for the asymptotic distribution
\eqref{eq:asymnormal}, forecast optimality relative to $\sigma(W)$
implies that
\[
J_T(\thh) \to \chi^2_{q-p} \; \text{ as } \; T \to \infty,
\]
where $J_T(\theta) = T \cdot g_T(\theta)' \, \Sigma_T^{-1}
g_T(\theta)$ is called the $J$-statistic.

An important aspect of our notion of optimality is that a point
forecast can only be defined as optimal with respect to a specific
functional and a specific \emph{information set}~\citep{Holzmann2014}.
The choice of the instrument vector $W$ determines the information set
for which we test.  If a forecast is optimal with respect to $\cF$, it
also satisfies the moment conditions for any information set $\cG
\subseteq \cF$.  In particular, if a test with instrument vector $W$
rejects optimality, the point forecast is deemed suboptimal with
respect to any information set $\cF$ that contains $\sigma(W)$.
Therefore, the null hypothesis in the test of overidentifying
restrictions is the existence of a parameter value $\thetatrue \in
\Theta$ such that $X$ is optimal at level $m(S,\thetatrue)$ relative
to information that includes the instrument vector $W$.

An optimal yet uninformed point forecast can only be rejected if
appropriate instruments are available.  Furthermore, a misspecified or
non-optimal forecast can still be optimal with respect to a smaller
information set or a more flexible class of functionals.  For these
reasons, the choice of the specification model and of a sufficiently
rich instrument vector are crucial.  By its very nature the current
forecast value carries substantial information about the underlying
quantile or expectile level, and we recommend strongly that it be
included as a component of the instrument vector.  The respective
effects are illustrated in the simulation setting of Section
\ref{sec:sim.statedependent} below.

The asymptotic results in Section \ref{sec:GMM} allow for the testing
of general hypotheses about forecasting behavior in customary ways.
For example, any restriction $R(\theta) = 0$ for the specification
model $m(s, \theta)$, where $R : \Theta \mapsto \real^l$ is
differentiable, can be tested for based on the Wald statistic
\citep[e.g.,][]{Greene2003}.

\section{Monte Carlo experiments}  \label{sec:simulation}

In this section we demonstrate that our GMM estimator is reliable in finite-sample settings, and induces well calibrated and powerful tests, despite its flexibility. For ease of comparison to the related approach of \citet{Patton2007},
which operates under the loss function paradigm, we adopt their
simulation setting.  Specifically, each sample path $y_1, \ldots, y_T$
is simulated from a stationary AR(1)--GARCH(1,1) model of the form
\begin{equation}  \label{eq:dgp}
	\textstyle
	Y_t  = \frac{1}{2} Y_{t-1} + \sigma_t \epsilon_t
	\quad \mbox{ where } \quad  
	\sigma^2_t = \frac{1}{10} + \frac{4}{5} \sigma^2_{t-1} + \frac{1}{10} \sigma^2_{t-1} \epsilon^2_{t-1},  
\end{equation}
with $\epsilon_t$ being standard Gaussian white noise.  Let $\cI_t$ be
the filtration generated by the time series, $\cI_t = \sigma(Y_t,
Y_{t-1}, \ldots)$.  We consider optimal forecasts based on distinct
information sets in Section \ref{sec:sim.info} and based on different
specification models in Section \ref{sec:sim.statedependent}. All
results use a Monte Carlo sample of 2000 replicate time series, and
all tests have nominal level 0.10.  Code for replication is available
at \url{https://github.com/Schmidtpk/pf}.

\subsection{State-independent forecasts under different information sets}  \label{sec:sim.info}

We follow \citet{Patton2007} and consider the point forecast 
\begin{equation}  \label{eq:dgp.x}
	\textstyle
	X_t = \frac{1}{2} Y_{t-1} - \frac{1}{4} \sigma_t.
\end{equation}
Conditional on $\cF_t = \cI_{t-1}$ the point forecast $X_t$ is an
optimal quantile-forecast at the constant level $\tau =
\Phi(-\tfrac{1}{4}) = 0.4012\ldots$ Alternatively, the forecast can be
interpreted as an optimal expectile-forecast at the constant level
$\tau = 0.3508\ldots$. The two interpretations are equally valid, as
the respective conditional distributions are all
Gaussian.\footnote{The same comment applies whenever the conditional
	distributions remain within a given location-scale family. See
	\citet[][Proposition 1]{Yao1996} and the discussion in Supplementary
	Sections \ref{supp:existence} and \ref{supp:statedependent}.}

\begin{figure}[t]
	
	\centering 
	\includegraphics[width=.75\textwidth]{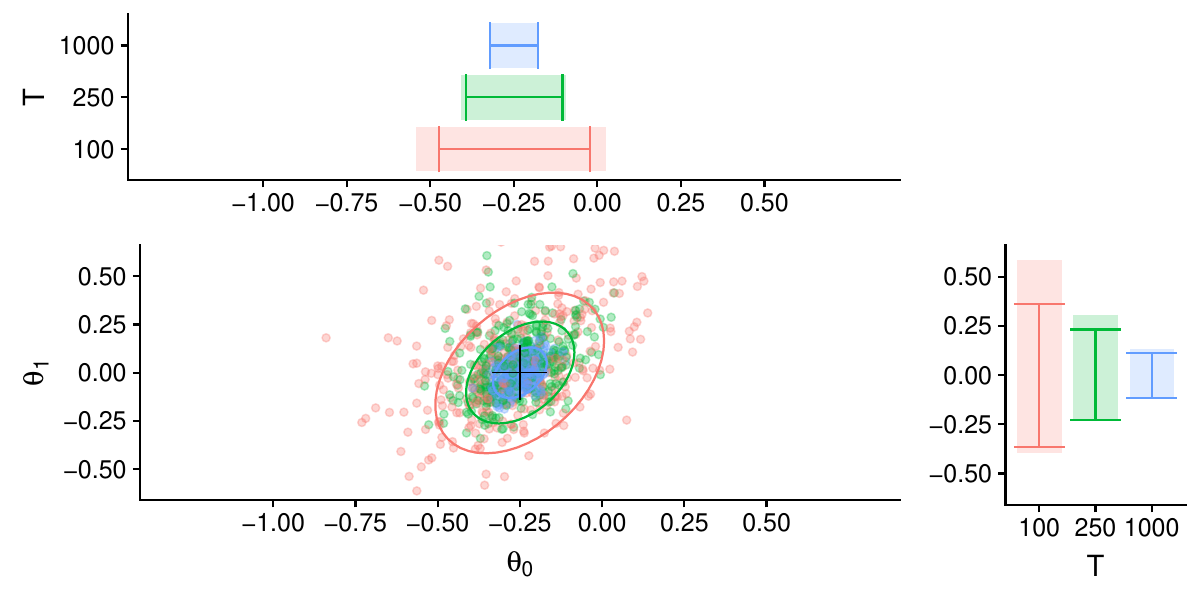} 
	
	\vspace{-2mm}
	
	\caption{Empirical distribution of the GMM estimate based on sample
		paths of size $T = 100, 250$, and $1000$ from \eqref{eq:dgp} and
		\eqref{eq:dgp.x}. The scatterplot shows estimates of $\theta =
		(\theta_0, \theta_1)'$ along with the respective 90\% ellipsoids
		from the large sample approximation. The true parameter values are
		at the center of the cross. The boxes at top and right range from
		the 5th to the 95th percentile of the estimates, as compared to the
		large sample approximation (bars).
		\label{fig:estimation}}
	
\end{figure}

Before discussing tests, we assess the finite-sample relevance of the
asymptotic distributions of the GMM estimator in Section
\ref{sec:GMM}. Specifically, consider the linear quantile
specification model with the point forecast $X_t$ as state variable.
The true parameter values in this setting are $\theta_0 = -0.25$ and
$\theta_1 = 0$.  As instruments we use a constant, the forecast, the
lagged forecast error, the squared lagged forecast error, and one
further lag of these variables, so that the instrument vector
\begin{equation}  \label{eq:IV} 
	W_t = \left( 1, X_t, X_{t-1} - Y_{t-1}, (X_{t-1} - Y_{t-1})^2, 
	X_{t-1}, X_{t-2} - Y_{t-2}, (X_{t-2} - Y_{t-2})^2 \right)'  
\end{equation}  
is of dimension $q = 7$.  Figure \ref{fig:estimation} illustrates the
empirical distribution of the GMM estimates based on sample paths of
size $T = 100$, 250, and 1000, respectively.  There is good agreement
with the large sample approximation of Section \ref{sec:GMM}.

\begin{figure}[t]
	
	\centering 
	\includegraphics[width=.55\textwidth]{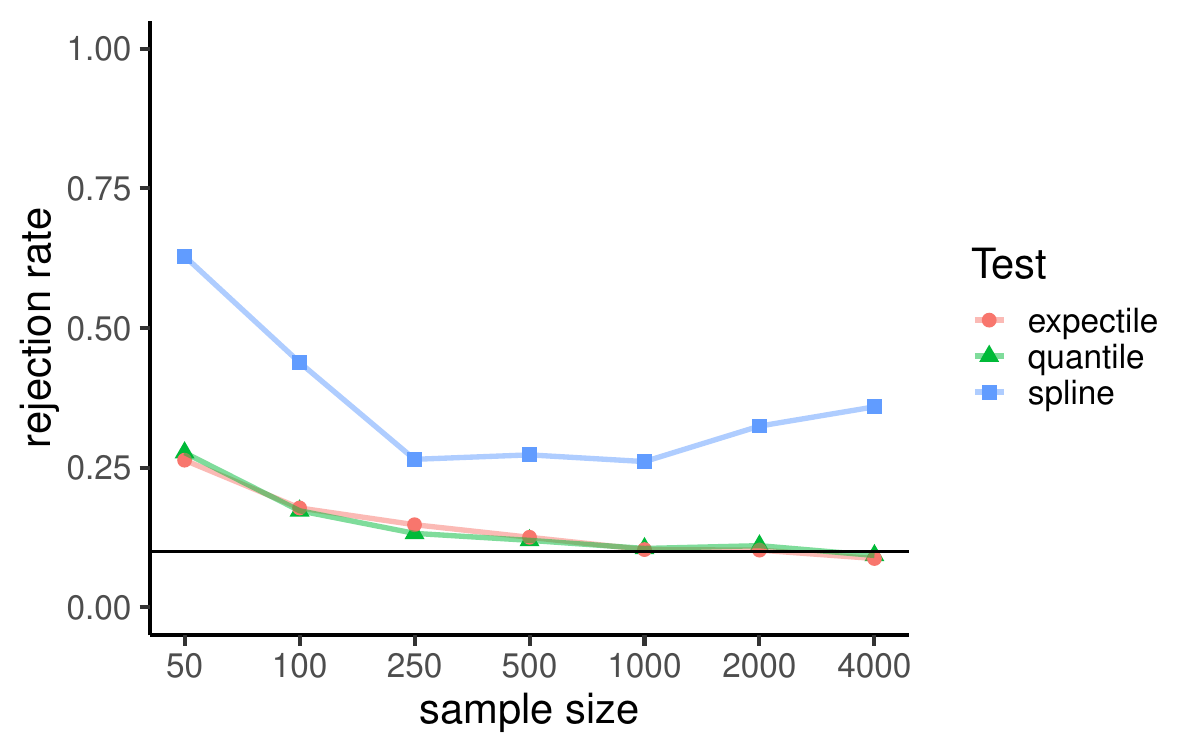}
	
	\caption{Size of optimality tests for the one-step ahead forecast.
		The horizontal line is at the nominal level of
		0.10.  \label{fig:sim.size}}
	
\end{figure}

Next we perform the overidentifying-restrictions test of forecast
optimality from Section \ref{sec:testing} with instrument vector $W_t$
in \eqref{eq:IV}.  We compare tests based on state-dependent quantiles
and expectiles with a linear specification model in the state variable
$X_t$ to the flexible spline test with state variable $Y_t$ as
specified in eq.~(16) of \citet{Patton2007}.  Figure
\ref{fig:sim.size} demonstrates that the expectile- and quantile-based
tests are better calibrated than the flexible spline test.  This
addresses a known problem of the state-dependent spline test, which
``appears to require large samples ($T \geq 1000$) before the test's
size is close to its nominal value, and thus rejections obtained using
this test must be interpreted with caution''
\citep[][p.~1183]{Patton2007}.

A possible reason is that, using a single node (at zero) only, the
flexible spline test reduces to the expectile test with a
linear-logistic specification model and an inadmissible (i.e., not
$\cI_{t-1}$-measurable) state variable $Y_t$, which seems problematic
from both theoretical and substantive points of view.  It is
interesting to observe that in the original loss function formulation
the spline approach seems entirely innocuous, with measurability
issues not being apparent at all.  Considering the spline with
admissible state variables, such as $Y_{t-1}$ or $X_t$, this critique
does not apply, but the general identifiability problem of the loss
function approach as described in Section \ref{sec:identification}
persists.  Indeed, in a range of simulation settings both with and
without state-dependence (in experiments not shown here) the
spline-based estimator appears to be unidentified and inconsistent.
In contrast, the approach based on state-dependent quantiles and
expectiles provides insightful point estimates, reliable confidence
intervals, and nearly nominally sized tests.

For the subsequent power analysis, we construct a two-step ahead
constant quantile- or expectile-forecast with respect to the lagged
information set $\cF_t = \cI_{t-2}$.  The respective conditional
predictive distributions are Gaussian with mean $\myE_{t-2}[Y_t] =
\tfrac{1}{4} Y_{t-2}$ and variance $\mbox{var}_{t-2}[Y_t] =
\tfrac{23}{20} \sigma_{t-1}^2 + \tfrac{1}{10}$, as shown in
Supplementary Section \ref{supp:derivation}.  Therefore,
\[
X_t =  \tfrac{1}{4} Y_{t-2} - \tfrac{1}{4} \sqrt{\tfrac{23}{20} \sigma_{t-1}^2 + \tfrac{1}{10}}  
\]
is an optimal two-step ahead quantile-forecast at level $\tau =
0.4012\ldots$, or an optimal expectile-forecast at the constant level
$\tau = 0.3508\ldots$, relative to the information set $\cI_{t-2}$ of
variables observable at time $t - 2$.  However, $X_t$ fails to be
optimal with respect to $\cI_{t-1}$.  This setting allows us to
evaluate the power of the optimality test against information
rigidities \citep{Coibion2015}.  A well performing test rejects the
optimality of $X_t$ based on information in $\cI_{t-1}$ but not
$\cI_{t-2}$, and obtains its nominal value with instruments in
$\cI_{t-2}$.  In the former case we use the instrument vector $W_t$ in
\eqref{eq:IV}, and in the latter case its lagged version $W_{t-1}$.

\begin{figure}
	
	\centering 
	\includegraphics[width=.55\textwidth]{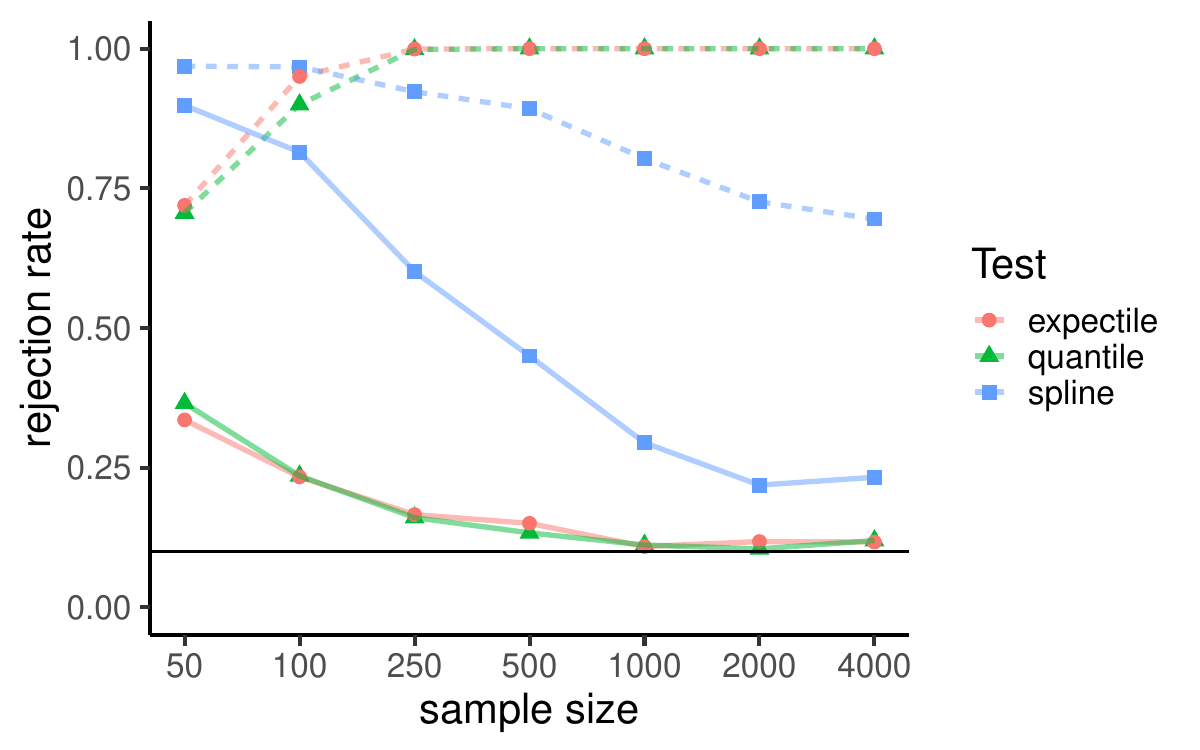}
	
	\caption{Size and power of optimality tests for the two-step ahead
		forecast.  The solid lines represent size for tests with properly
		lagged instruments.  The dashed lines represent power for tests with
		non-lagged instruments.  \label{fig:sim.power}}
	
\end{figure}

We employ quantile-, expectile-, and spline-based tests with the same
specifications as before.  The results of this experiment are
presented in Figure \ref{fig:sim.power}.  The expectile- and
quantile-based optimality tests are better calibrated and more
powerful than the spline-based test, which is strongly oversized for
small sample sizes and unable to consistently detect the information
rigidity even for large sample sizes.  Hence, our tests are not only
better calibrated, but also more powerful.  While for the sake of
comparability we have kept the instrument vector identical across
tests in this experiment, the functional-based tests benefit further
under more flexible models, where the heavily parameterized spline
estimator requires instrument vectors of considerably higher dimension
than in the quantile- and expectile-based approaches.

\subsection{State-dependent forecasts under different specification models} 
\label{sec:sim.statedependent}

\begin{figure}
	
	\medskip
	
	\centering 
	\includegraphics[width=.65\textwidth]{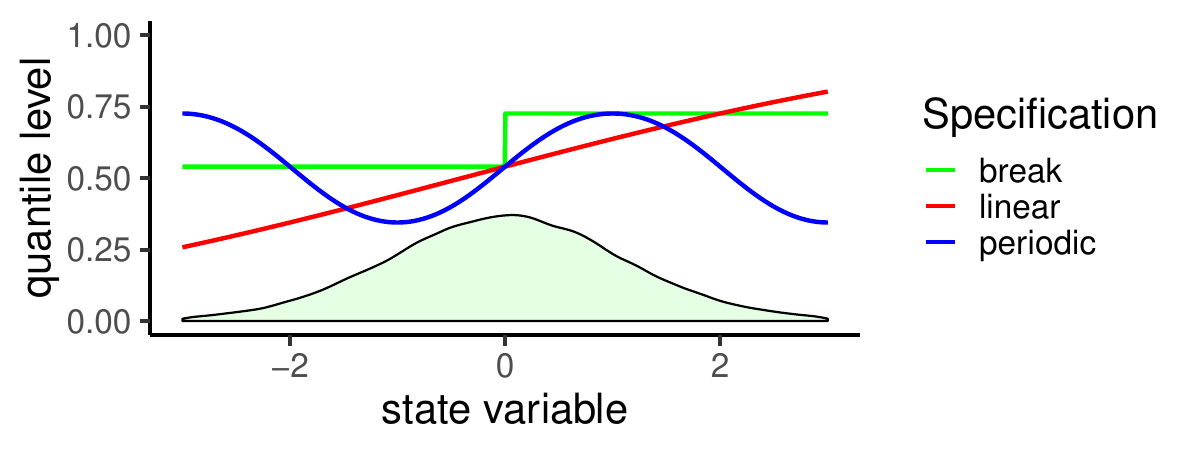}
	
	\caption{Quantile level as a function of the state variable under the
		specification models in Section \ref{sec:sim.statedependent}.  The
		density of the state variable $Y_{t-1}$ is also
		shown.  \label{fig:specifications}}
	
\end{figure}

Next we investigate whether our optimality test can discriminate among
the proposed specification models.  To this end, we generate optimal
state-dependent forecasts for the data generating process
\eqref{eq:dgp} with the most recent outcome of the time series,
$Y_{t-1}$, as state variable.  Specifically, we let
\[
X_t =  \tfrac{1}{2} Y_{t-1} + q_{m(Y_{t-1})}(\cN(0,1)) \sigma_t,  
\]
where $m(s) = \Phi(\tfrac{1}{10} + \tfrac{s}{4})$ under the linear
specification model, $m(s) = \Phi( \tfrac{1}{10} + \tfrac{1}{2} \cdot
\one{ s \geq 0})$ under the break model, and $m(s) =
\Phi(\tfrac{1}{10} + \tfrac{1}{2} \sin(\frac{\pi}{2} s))$ under
the periodic model, as illustrated in Figure \ref{fig:specifications}.
We consider specification models with two parameters only; to achieve
this, we fix the break point and the period in the break and periodic
model at their respective true values.

We then apply overidentifying-restrictions tests of forecast
optimality with instrument vector $W_t = (1, Y_{t-1}, X_t)'$.  Table
\ref{tab:sim.reject} shows results for forecasts and tests based on
quantiles.  Even for small sample sizes our tests are reasonably
calibrated and quite powerful with rejection rates up to 70\% for more
distinct specifications.  For larger sample sizes the optimality tests
are almost perfectly calibrated with high power.

\begin{table}
	
	\caption{Rejection rates of optimality tests based on quantile
		specification models with state variable $Y_{t-1}$.  The nominal
		level is 0.10, and the GMM estimator uses the instrument vector $W_t
		= (1, Y_{t-1}, X_t)'$ or $W_t = (1, Y_{t-1}, Y_{t-2})'$,
		respectively.  Settings where the null hypothesis is satisfied are
		marked in gray.
		\label{tab:sim.reject}}
	
	\begin{tabular}{c|l|ccc|ccc}
		& true model & \multicolumn{6}{c}{hypothesized model} \\
		\hline
		sample & quantile & \multicolumn{3}{c|}{$W_t = (1, Y_{t-1}, X_t)'$} 
		& \multicolumn{3}{c}{$W_t = (1, Y_{t-1}, Y_{t-2})'$}  \\
		size   &          & linear & break & periodic & linear & break & periodic \\
		\hline
		& linear   & \mycellcolor 0.06 & 0.14 & 0.31 & \mycellcolor 0.11 & 0.11 & 0.10 \\ 
		$T=100$   & break    & 0.32 & \mycellcolor 0.11 & 0.09 & 0.11 & \mycellcolor 0.10 & 0.10 \\  
		& periodic & 0.68 & 0.33 & \mycellcolor 0.08 & 0.11 & 0.11 & \mycellcolor 0.10 \\  
		\hline
		& linear   & \mycellcolor 0.09 & 0.35 & 0.84 & \mycellcolor 0.10 & 0.10 & 0.10 \\ 
		$T = 250$ & break    & 0.55 & \mycellcolor 0.09 & 0.12 & 0.11 & \mycellcolor 0.10 & 0.10 \\ 
		& periodic & 0.97 & 0.63 & \mycellcolor 0.07 & 0.10 & 0.09 & \mycellcolor 0.10 \\
		\hline
		& linear   & \mycellcolor 0.11 & 0.84 & 1.00 & \mycellcolor 0.09 & 0.10 & 0.13 \\ 
		$T = 1000$ & break    & 0.96 & \mycellcolor 0.10 & 0.39 & 0.10 & \mycellcolor 0.10 & 0.10 \\ 
		& periodic & 1.00 & 0.99 & \mycellcolor 0.09 & 0.10 & 0.10 & \mycellcolor 0.10 \\ 
		\hline
	\end{tabular} 
	
\end{table}

As noted, the power of the tests depends crucially on the choice of
the instrument vector.  The final block of columns in Table
\ref{tab:sim.reject} considers the same setting as before, except for
the instruments used.  Specifically, we drop the forecast value, and
now use a constant and the first two lagged outcomes of the time
series, so that $W_t = (1, Y_{t-1}, Y_{t-2})'$.  While the tests
continue to be well calibrated their power is diminished.
Supplementary Section \ref{supp:stateX} considers the case where the
state variable in the linear quantile specification model is the point
forecast $X_t$, rather than the most recent outcome, and Supplementary
Section \ref{supp:statedependent} provides further results in the case
where expectile specification models are hypothesized in lieu of the
data-generating quantile models, with similar findings.  We therefore
recommend that the forecast value, which by its very nature carries
information about the underlying forecast directive, be included in
the instrument vector.

\section{Empirical examples}  \label{sec:empirical} 

Unless noted otherwise, we apply the linear specification model with
the probit link function and the current forecast value as state
variable, and the instrument vector includes a constant, the forecast
value at hand, and the most recent outcome available when the forecast
was issued.

\subsection{Gross domestic product (GDP) growth forecasts as state-dependent quantiles}  \label{sec:Greenbook}

We revisit the well studied Greenbook forecasts of the US~Federal
Reserve for GDP growth.  As multiple forecasts are issued within a
quarter for the GDP growth in the next quarter, we consider two
different one-quarter ahead forecasts. Specifically, for each GDP
observation we consider the forecast issued closest to the midpoint
(main forecast) and closest to the end (late forecast) of the previous
quarter.  As realized values we take the quarterly real GDP growth
rate in the US over the period 1969 to 2012 for $T = 176$
observations, as reported in the initial data release and illustrated
in Figure \ref{fig:data_gdp}.\footnote{The results in this section are
	robust to using the second revision or the most recent vintage.}  In
a pioneering effort, \citet{Patton2007} modeled the Federal Reserve's
loss function as a quadratic spline with three nodes whose shape is
allowed to change with the realized growth rate.

Here we interpret the forecasts as quantiles of the Federal Reserve's
(implicit) predictive distributions. As instrumental variables for the
GMM estimator we employ a constant, the one-step ahead forecast at
hand, and the most recent outcome available at the time of the
forecast, i.e., $w_t = (1, y_{t-2}, x_t)'$.  For the late forecast
issued at the end of the previous quarter, there is no evidence
against the hypothesis of an optimal quantile forecast at a constant
level.  The estimated quantile level is 0.59 with standard error 0.04,
and the $p$-value in the associated test of overidentifying
restrictions is 0.29.

For the main forecast the hypothesis of an optimal quantile forecast
at a constant level is untenable, as the $p$-value in the test of
overidentifying restrictions drops to 0.05.  To investigate whether
the main forecast is optimal if we allow the reported quantiles to
change with the predicted GDP growth rate, $x_t$, we apply the linear
specification model $m(x_t, \theta) = \Phi(\theta_0 + \theta_1
x_t)$. Compared to the spline loss function of \citet{Patton2007}, we
use two only instead of six parameters, and we apply state variables
and instruments that are at least implicitly available at the time
when the forecast is issued.  The test of overidentifying restrictions
yields a $p$-value of $0.49$.

\begin{figure}
	
	\centering
	
	\includegraphics[width=.55\textwidth]{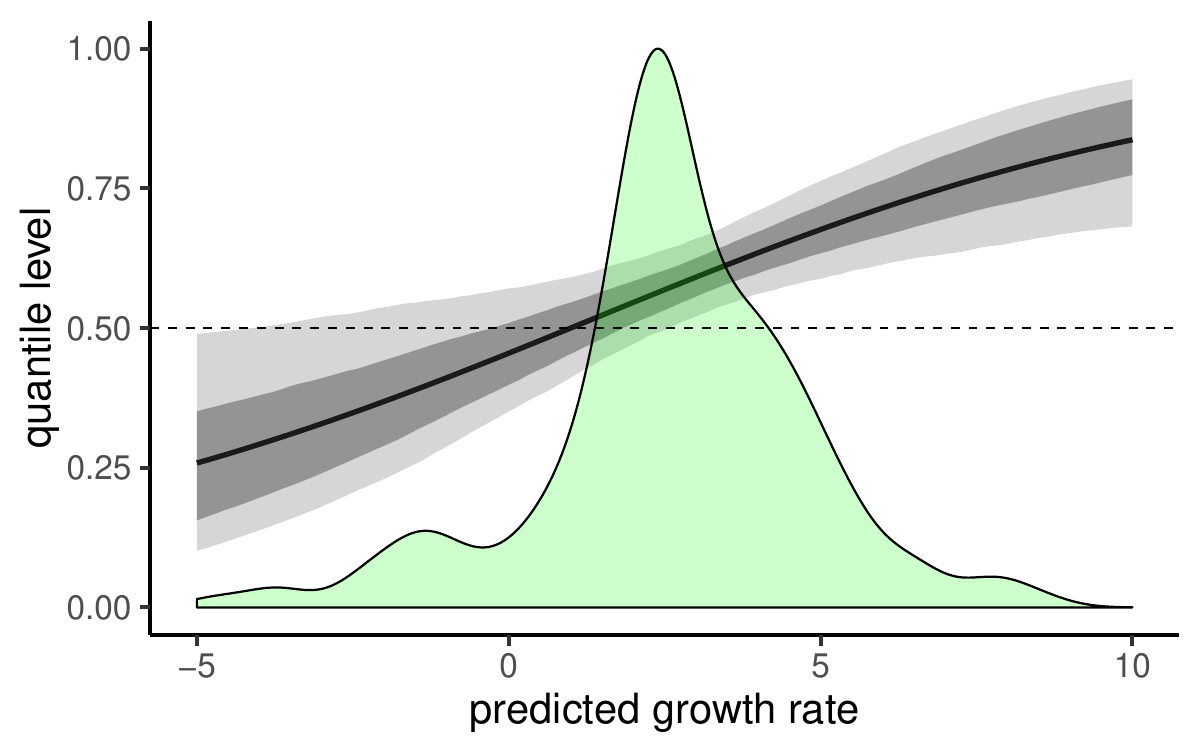}
	
	\caption{Estimated quantile specification model for the Federal
		Reserve's forecasts of GDP growth plotted against the predicted
		growth rate (in percent), with pointwise confidence intervals at
		level 0.60 and 0.90.  A density estimate of the point forecast is
		also shown.  \label{fig:Greenbook}}
	
\end{figure}

The GMM estimate for $\theta = (\theta_0, \theta_1)'$ is $\thh =
(-0.10, 0.11)'$, and we interpret the forecasts as
$m(x_t,\thh)$-quantiles that depend on the predicted growth rate
$x_t$.  The covariance estimate implied by \eqref{eq:asymnormal} is
given by
\[
\frac{1}{T} \left( \hat{G}_T' \, \hat{\Sigma}^{-1}_T \hat{G}_T \right)^{-1} 
=
\begin{pmatrix}
0.028 & -0.006 \\
-0.006 & 0.002
\end{pmatrix},
\]
where $\hat{G}_T$ is the sample moment of $G$ evaluated at $\theta =
\thh$.  As $\Phi$ is strictly monotone, we can compute pointwise
confidence intervals for $\hth_0 + \hth_1 x_t$ and transform into
confidence intervals for $m(x_t,\hth) = \Phi(\hth_0 + \hth_1 x_t)$, as
illustrated in Figure \ref{fig:Greenbook}.  The Federal Reserve
reports higher quantile levels during times of strongly positive
expected growth.  The model also stipulates lower quantile levels in
times of negative expected growth, but the confidence bands do not
exclude the median forecast except for extreme cases.

We summarize that it is key to consider the available information in
tests of forecast optimality.  While the late Greenbook forecasts
issued toward the end of the quarter are optimal in the classical
sense, the driving factor of overly optimistic forecasting earlier in
the quarter is anticipated growth, whereas the behavior in
anticipation of recessions is inconclusive.

\subsection{Precipitation forecasts as state-dependent expectiles}  \label{sec:ECMWF}

\begin{figure}
	
	\centerline{\includegraphics[width=.55\textwidth]{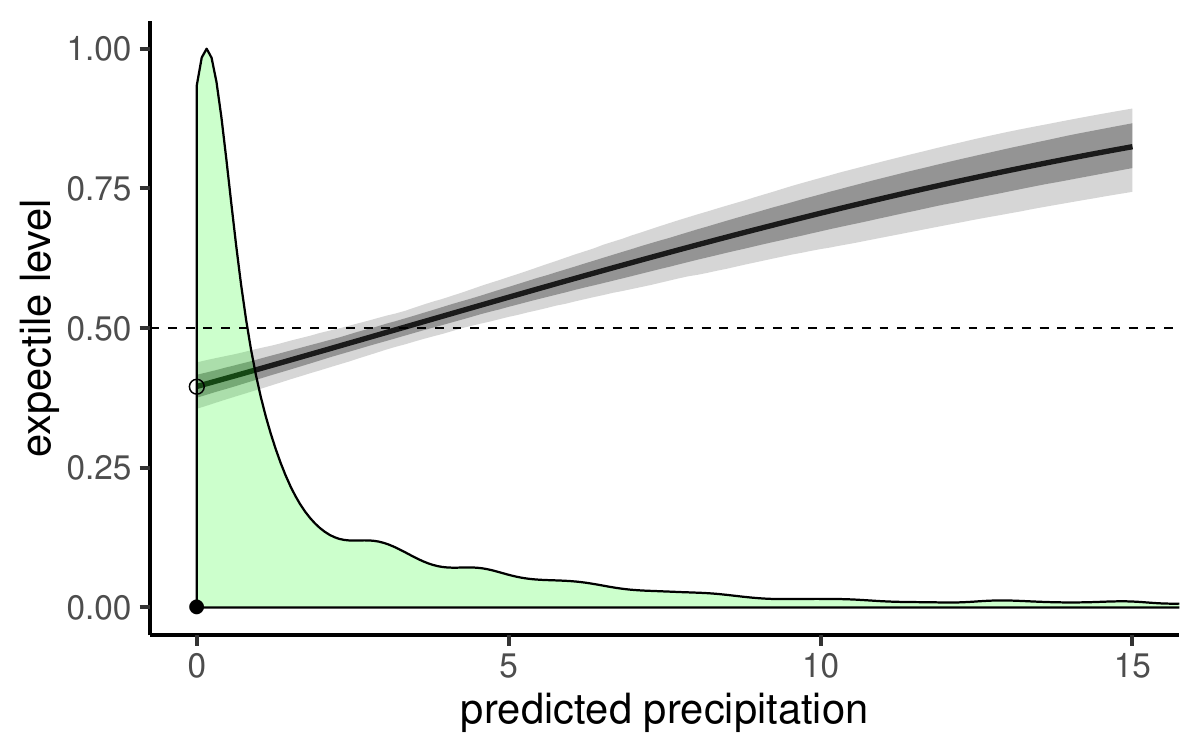}}
	
	\caption{Estimated expectile specification model for ECMWF forecasts
		of daily rainfall at London plotted against the predicted
		precipitation level (in mm), with pointwise confidence intervals at
		level 0.60 and 0.90.  A density estimate of the point forecast
		conditional on it being strictly positive is also
		shown.  \label{fig:ECMWF}}
	
\end{figure}

Numerical weather prediction has seen tremendous advances over the
past decades \citep{Bauer2015}. Here we consider 24-h ahead forecasts
of daily precipitation accumulation over London, UK from the
high-resolution run operated by the European Centre for Medium-Range
Weather Forecasts (ECMWF; \url{www.ecmwf.int}).  While the forecasts
are generated by numerical models that are run on supercomputers
\citep{ECMWF2012}, many facets of the operational implementation are
subject to tuning by human experts, often based on decade-long
experience.  In our analysis we use ECMWF forecasts and observations
from the ERA-Interim system \citep{ERA-Interim} in the period 1 July
2011 to 30 June 2017 for $T = 2192$ observations, as partly
illustrated in Figure \ref{fig:data}.  Despite good agreement between
forecasts and observations, the hypothesis of an optimal constant
expectile forecast is rejected with a $p$-value $< 10^{-8}$.  Instead, 
we interpret the forecasts as state-dependent expectiles of the
underlying predictive distribution.  We employ expectiles rather than
quantiles to avoid artifacts due to the mixed discrete-continuous
nature of precipitation accumulation, which is a nonnegative variable
with a point mass at zero.  Any point forecast $x_t = 0$ can be
interpreted as essential infimum or expectile at level $\tau = 0$.  To
investigate whether the reported expectile varies with predicted
precipitation accumulations $x_t > 0$, we apply a slightly modified
linear specification model,
\[
m(x_t, \theta) = \Phi(\theta_0 + \theta_1 x_t) \, \one{x_t > 0}.
\]
The GMM estimate with instrument vector $w_t = (1, y_{t-2}, x_t)'$ for
$\theta = (\theta_0, \theta_1)'$ is $\thh =
(-0.27,0.08)'$.\footnote{These results are robust to using the
	instrument vector $w_t = (1, y_{t-1}, x_t)'$. However, as $y_{t-1}$
	represents the daily precipitation accumulation, it has not yet been
	observed when the numerical model that generates $x_t$ is
	initialized.}  The resulting specification model is illustrated in
Figure \ref{fig:ECMWF}, and the test of overidentifying restrictions
does not reject forecast optimality, with the $p$-value being at 0.74.
It is interesting to observe that the numerical model generates
considerably higher expectile levels in anticipation of severe,
extreme rain.

\section{Discussion}  \label{sec:discussion}

For point forecasts with unknown directive, we posit that it is
preferable to estimate and test the functional quoted by the
forecaster, rather than the loss function, for reasons of
identifiability, interpretability, and ease and efficiency of
inference.  While our approach deviates from recent literature, the
classical \citet{Mincer1969} test of forecast optimality can be
interpreted in our setting.\footnote{\citet{Mincer1969} employ the
	regression model $y_t = \beta_0 + \beta_1 x_t + u_t$ and test
	whether the coefficients equal zero and one respectively.  This can
	be interpreted as assuming an optimal forecast of the functional
	form $\beta_0 + \beta_1 \myE[Y|X]$ with identification function
	$V(x,y) = \beta_0 + \beta_1 x - y$, and applying the GMM estimator
	with instruments $w = (1,x)'$, with ensuing tests.}

We have introduced quantile and expectile specification models for the
description of forecasting behavior in terms of a state variable,
thereby relating to extant work on time-varying quantiles and
expectiles \citep{DeRossi2009}.  Under the assumption of optimal
forecasts, the model parameters can be consistently estimated, and the
asymptotic distributions of the GMM estimator and the respective test
statistics can be used to construct flexible tests of forecast
optimality and specific model properties.  It is particularly
noteworthy that state-dependent functionals allow for the treatment of
supposedly misspecified forecasts in a principled manner.
Furthermore, support conditions and mixed discrete-continuous
distributions can be handled efficiently and with rigour, as
illustrated in the precipitation example.

For valid estimation and testing, the state variables need to be in
the information set of the forecaster. A valid and universally
available state variable is the forecast value at hand.  The GMM
estimator depends on an appropriate choice of a sufficiently rich
instrument vector, and we recommend that the forecast value at hand be
included as an instrumental variable. Importantly, the null hypothesis
in the overidentifying-restrictions test of forecast optimality
reflects the selection of the instrument vector, and judicious choices
are critical, as we have demonstrated in real data and simulation
examples.  An accompanying software package in {\sc R} \citep{R} is
available at \url{https://github.com/Schmidtpk/PointFore}.

Our results rely on the assumption of a general version of forecast
optimality, or more specifically on a parametric relation between the
point forecast and the ideal predictive distribution.  In a more
elaborate approach, this assumption is substituted for by assumptions
on the forecast generation, e.g., via a correctly specified time
series model and efficient estimation in an increasing training
window, thereby accounting for estimation uncertainty
\citep{Elliott2005, Guler2017}.  Reassuringly, in the case of a
constant specification model, our asymptotic results are identical to
extant results that account for estimation uncertainty
\citep[][Proposition 4]{Elliott2005}.  However, in unstable
environments, our optimality assumption may be too strong and a more
elaborate description that accounts for estimation uncertainty may be
warranted.

In empirical examples from economics and meteorology, we have applied
the linear specification model with the forecast at hand as state
variable, and we have found common ground, in that the forecasts are
indicative of overstating in anticipation of extreme events.  While
application-specific reasons for this remain to be explored, a
potential interpretation is via the forecasters's dilemma
\citep{Lerch2017}, which refers to the fact that the public's
attention focuses on the forecast performance in cases of extreme
events.  As a result of this ubiquitous practice, individual and
institutional forecasters may have implicit incentives to exaggerate
in anticipation of tail events.  A related phenomenon is the hard-easy
effect described in the psychological literature, in that human
subjects tend to be overconfident in answering hard questions, while
being underconfident in responses to easy questions
\citep{Lichtenstein1982, Kynn2008}.

Further investigation is called for in order to assess whether
overstatement in anticipitation of extreme events is characteristic
when predictions are generated by individual or institutional
forecasters, or when numerical models are designed, tuned, and
informed by human expertise.  The commonalities between our two unlike
examples from economics and meteorology suggest that the linear
specification model with the forecast value as state variable might be
a useful default choice in studying this type of question.

\subsection*{Acknowledgments}

We are grateful to the editor, the associate editor, two anonymous
referees, Werner Ehm, Alexander Glas, Fabian Kr\"{u}ger, Barbara Rossi
and Peter Vogel for a wealth of constructive and insightful comments.
Furthermore, we thank the European Centre for Medium-Range Weather
Forecasts (ECMWF) in Reading, UK for providing meteorological data,
and we are grateful to Stephan Hemri for assistance in their handling.
The work of Patrick Schmidt, Tilmann Gneiting and Stephan Hemri was
partially funded by the Klaus Tschira Foundation and by the European
Union Seventh Framework Programme under grant agreement no.~290976.
Tilmann Gneiting also is grateful for travel support and encouragement
through the ECMWF Fellowship programme.  Matthias Katzfuss was
partially supported by US~National Science Foundation (NSF) Grant
DMS--1521676 and NSF CAREER Grant DMS--1654083.

\small
\appendix

\section*{Appendix}

\section{Identifying moment conditions}  \label{app:ident}

Consider the probability space $(\Omega, \cA, \myP)$, where the
elements of the sample space $\Omega$ are tuples that comprise the
point forecast $X$, the realization $Y$, and a covariate vector $Z$.
We assume that the information set $\cF$ is a sub-$\sigma$-algebra of
$\cA$.  If no measure is explicitly mentioned, statements like almost
surely refer to $\myP$.  For random variables $R_1$ and $R_2$, we
simply write $R_1 = R_2$ instead of $R_1 = R_2$ almost surely.  In
particular, statements like $X = \alpha(Y|\cF)$ denote $\myP$-almost
sure properties.  As defined in Section \ref{sec:identification}
standard measurability and integrability conditions are denoted $R \in
\cF_Q$ and $R \in \cF$.

Before proceeding to our main results on identifying moment
conditions, we state an elementary measure-theoretic equivalence.

\begin{lemma}  \label{le:condexp}
	For every integrable random variable $U$, 
	\[
	\myE(U|\cF) = 0 \iff 
	\myE [UW] = 0 \; \text{ for all } \; W \in \cF_U.
	\]
\end{lemma}

\begin{proof}	
	The implication from left to right is immediate from Theorem 34.3 in
	\citet{Billingsley}.  For the reverse implication let $W$ be the
	indicator function of any event $A$ in the information set $\cF$, to
	yield $\int_A \myE(U|\cF) \dd \myP = 0$ for all $A \in \cF$, which
	implies $\myE(U|\cF) = 0$ by a standard argument.
\end{proof}

We now consider the $\tau$-quantile functional $q_\tau$ and the
$\tau$-expectile functional $e_\tau$, which includes the special case
$\tau = \frac{1}{2}$ of the mean-functional.  The following
assumptions ensure that the functional is single-valued and well
defined.

\begin{enumerate}
	
	\item[A$_Q$] The conditional distribution $\cL(Y|\cF)$ is absolutely
	continuous with a strictly increasing cumulative distribution
	function almost surely.
	
	\item[A$_E$] The conditional distribution $\cL(Y|\cF)$ has finite mean
	and positive variance almost surely.
	
\end{enumerate}

\begin{lemma}[quantiles]  \label{le:ident.q}
	Under condition A$_Q$ the function $V_\tau(x,y) = \one{x \geq y} -
	\tau$ identifies the optimal $\tau$-quantile forecast, i.e.,
	\[
	X = q_\tau(Y|\cF) \iff \myE [(\one{X \geq Y} - \tau) W] = 0 
	\; \text{ for all } \; W \in \cF.
	\]
\end{lemma}

\begin{proof}	
	For every $\omega \in \Omega$ the definition of the $\tau$-quantile
	implies that $x = q_\tau(Y|\cF)(\omega) \iff \myE [V_\tau(x,Z)] = 0$,
	where $Z \sim \cL(Y|\cF)(\omega) \in \cP$.  In terms of the
	$\cF$-measurable random variable $X$ this equality can be stated as $X
	= q_\tau(Y|\cF) \iff \myE [V_\tau(X,Y)|\cF] = 0$.  The stated
	equivalence is now immediate from Lemma \ref{le:condexp}.
\end{proof}

\begin{lemma}[expectiles]  \label{le:ident.e}
	Under condition A$_E$ the function $V_\tau(x,y) = |\one{x \geq y} -
	\tau| (x-y)$ identifies the optimal $\tau$-expectile forecast, i.e.,
	\[
	X = e_\tau(Y|\cF) \iff \myE [|\one{X \geq Y} - \tau| (X-Y) W] = 0 
	\; \text{ for all } \; W \in \cF_{Y-X}.
	\]
\end{lemma}

The proof is essentially the same as in the case of quantiles.  The
next and final result in this section is a variant of findings in
\citet{Steinwart2014}, and we follow the terminology used in their
paper.  In particular, topological statements on the space of
probability distributions with bounded Lebesgue measures are with
respect to the metric induced by the $L_1$-norm.  The conditions on
the functional $\alpha$ are met if it is defined via a continuous,
non-trivial loss function, for continuity follows from the Maximum
Theorem \citep[e.g.,][p.~229]{Ok2007}, and functionals defined via
loss functions have convex level sets \citep{Osband1985,
	Gneiting2011}.

\begin{lemma}  \label{le:ident}
	Let $\cP$ be a convex set of probability measures with bounded
	Lebesgue densities such that $\cL(Y|\cF) \in \cP$ almost surely, and
	suppose that the functional $\alpha : \cP \mapsto \real$ is continuous
	and locally nonconstant with convex level sets.  Then there exists a
	measurable function $V_\alpha$ that identifies the optimal
	$\alpha$-forecast, i.e.,
	\[
	X = \alpha(Y|\cF) \iff \myE [V_\alpha(X,Y) W] = 0 \; \text{ for all } \;
	W \in \cF_{V_\alpha(X,Y)}.
	\]
\end{lemma}

\begin{proof}	
	By Theorem 8 in \citet{Steinwart2014} there exists a function
	$V_\alpha$ such that for all $P \in \cP$ it holds that $t = \alpha(P)
	\iff \myE_{Y \sim P}[V_\alpha(t,Y)] = 0$.  Using the same arguments as
	in the proof of Lemma \ref{le:ident.q}, we see that
	\[
	X = \alpha(Y|\cF) \iff \myE[V_\alpha(X,Y)|\cF] = 0, 
	\]
	and an application of Lemma \ref{le:condexp} completes the proof.
\end{proof}

\section{Consistency of the GMM estimator}  \label{app:consistency}

In order to establish consistency for the GMM estimator \eqref{eq:GMM}
in Section \ref{sec:estimation}, we extend the probability space
$(\Omega, \cA, \myP)$ to the dynamic prediction space setting of
\citet{Strahl2017} and apply classical GMM theory \citep{Hansen1982}.
As noted, statements about all time points $t \in \zz$ are typically
written without subscripts.  We define $u = (x,y,s)$ and denote the
identification function by $V(u,\theta)$, where $V(u,\theta) = \one{y
	\leq x} - m(s,\theta)$ in the case of quantiles and $V(u,\theta) =
|\one{y \leq x} - m(s,\theta)| (x-y)$ in the case of expectiles.

We employ the following assumptions.

\begin{enumerate}
	
	\item[B$_1$] The stochastic process $(U_t,W_t)$ is strictly stationary and
	ergodic.
	
	\item[B$_{2,Q}$] The components of the instrument vector $W$ have
	finite first moment.
	
	\item[B$_{2,E}$] The components of the vector $(X - Y) W$ have finite
	first moment.
	
	\item[B$_3$] The parameter space $\Theta \subseteq \real^p$ is
	compact, and the specification model $m(s,\theta)$ is continuous on
	$\Theta$ for all $s$ and Borel measurable for each $\theta \in \Theta$.
	
	\item[B$_4$] The specification model is uniquely identified by the
	instrument vector $W$, i.e., 
	\[
	\myE [V(U,\theta) W] = 0 \iff \theta = \thetatrue.
	\]
	
	\item[B$_5$] The weighting matrix $M_T$ converges almost surely to a
	constant matrix with full rank.
	
\end{enumerate}

Concerning the unique identification condition B$_4$, general insights
apply, in that a specification model with $p$ parameters calls for an
instrument vector of dimension $q \geq p$.  Consider for example a
univariate model $m(s, \theta)$ that is strictly monotone in $\theta$
for all $s$.  As the identification functions of quantiles and
expectiles are oriented, the trivial (constant) instrument suffices
for unique identification; cf.~\citet[Proposition
1]{Elliott2005}.  

For another example, let 
\[
m(s,\theta) = \theta' s = \sum_{i=1}^p \theta_i s_i
\]
for $s = (s_1, \ldots, s_p)'$ with components that are pairwise
uncorrelated and have positive variances.  This na\"{i}ve linear
specification model operates under severe constraints on the domain of
the parameters and the state variables, but is attractive from the
perspective of interpretation.  Under the quantile model the
identification function satisfies $V(u,\theta) = \one{y \leq x} -
m(s,\theta)$ and so it holds that $\myE [V(U,\theta) | \cF] =
(\thetatrue - \theta)' S$ and, consequently,
\[
\myE[V(U,\theta)W] = 0 \iff \left( \thetatrue - \theta \right)' 
\myE \! \left[ \left( \sum_{i=1}^p S_i \! \right) \! W \right] = 0. 
\]
As $\myE[S_i^2] + \sum_{j=1, j \neq i}^k \myE[S_i S_j] = \myE[S_i^2] >
0$ for $i = 1, \ldots, p$, condition B$_4$ is satisfied when $W = S$.

Generally, a specification model is uniquely identified if the
instrument vector $W$ generates the information set $\mathcal{F}$ and
$m(S,\theta) = m(S,\thetatrue)$ only if $\theta = \thetatrue$.  To see
this, note that
\[
\myE[(V(U,\theta) - V(U,\thetatrue)) W] 
= \myE[ \, \myE[V(U,\theta) - V(U,\thetatrue) | \cF] \, W] 
= 0
\]
in concert with $\sigma(W) = \cF$ implies $\myE[V(U,\theta) -
V(U,\thetatrue)) | \cF] = 0$.  For quantile models it holds that
$\myE[V(U,\theta) - V(U,\thetatrue) | \cF] = m(S,\theta) -
m(S,\thetatrue)$, so $\theta = \thetatrue$, which implies B$_4$.  For
expectile models $\myE[V(U,\theta) - V(U,\thetatrue) | \cF] =
(m(S,\theta) - m(S,\thetatrue)) \, \myE[X-Y | \cF]$, whence $\theta =
\thetatrue$ under the further condition that $X$ is not the optimal
mean-forecast.

For a first step estimator with the identity matrix as weighting
matrix condition B$_5$ is trivially satisfied.  Corollary
\ref{cor:HAC} in Appendix \ref{app:asnormal} provides sufficient
conditions for the consistency of the HAC estimator \citep{Newey1987a}
of the covariance matrix of the moment function, which implies B$_5$
for a sub-sequence of the inverse of the HAC matrix in the efficient
two-step procedure.

\begin{theorem}[Consistency]  \label{thm:consistency}
	If\/ $X_t$ is an optimal state-dependent quantile, i.e., $$X_t =
	q_{m(S_t,\thetatrue)}(Y_t|\cF_t)$$ with some $\cF_t$-measurable state
	variable $S_t$, conditions A$_Q$, B$_1$, B$_{2,Q}$, B$_3$, B$_4$, and
	B$_5$ guarantee existence and almost sure convergence of the GMM
	estimator.  Analogously, if\/ $X_t$ is an optimal state-dependent
	expectile, conditions A$_E$, B$_1$, B$_{2,E}$, B$_3$, B$_4$, and B$_5$
	yield existence and almost sure convergence.
\end{theorem} 

\begin{proof} 
	It suffices to verify the conditions of Theorem 2.1 in
	\citet{Hansen1982}, namely, Assumptions 2.1--2.5, (i), (ii), and
	(iii).  Assumption 2.1 follows from B$_1$ as $g(\theta) = V(u, \theta)
	w$ is a function of finitely many, jointly stationary and ergodic
	variables.  Assumptions 2.2 and 2.3 are immediate from B$_3$,
	Assumption 2.4 is guaranteed by B$_{2,Q}$ or B$_{2,E}$ along with
	B$_4$ as $|g(\theta)|$ is bounded by $|w|$ in the case of quantiles
	and $|(x-y)w|$ in the case of expectiles, and Assumption 2.5 follows
	from B$_5$.  Finally, an application of Lemma 2.1 in
	\citet{Hansen1982} under B$_{2,Q}$ or B$_{2,E}$ establishes (i), B$_3$
	yields (ii), and B$_4$ and B$_5$ guarantee (iii).
\end{proof}

\section{Asymptotic normality of the GMM estimator}  \label{app:asnormal}

Drawing again on well established GMM theory \citep{Hansen1982}, we
proceed to state sufficient conditions for consistent covariance
estimation and asymptotic normality of the GMM estimator.  As before,
we consider the function $g(\theta) = V(u, \theta) w$ as a mapping
from $\Theta \subseteq \real^p$ into $\real^q$.  Consistency is now
understood in the sense of convergence in probability.

\begin{enumerate}
	
	\item[C$_1$] The stochastic process $(U_t,W_t)$ is strictly
	stationary with mixing coefficients $\alpha_m$ of order ${\cal
		O}(m^{-s})$ for some $s > 2$.
	
	\item[C$_{2,Q}$] There exists $\delta > 0$ such that the components of the instrument 
	vector $W$ have finite absolute moment of order $4 + \delta$.
	
	\item[C$_{2,E}$] There exists $\delta > 0$ such that the components of
	the vector $(X-Y)W$ have finite absolute moment of order $4 +
	\delta$.
	
	\item[C$_3$] The true parameter value $\thetatrue$ is in the interior of
	$\Theta$.
	
	\item[C$_4$] The derivative\/ $m_{(\theta)}(s,\theta)$ exists is
	bounded and locally Lipschitz continuous at $\thetatrue$ uniformly in
	$s$, i.e. there exists $\delta>0$ such that $|m_{(\theta)}(s,\theta)
	- m_{(\theta)}(s,\thetatrue)| \leq K |\theta - \thetatrue|$ for all $s$
	and all $\theta$ with $|\theta - \thetatrue| < \delta$.
	
	\item[C$_5$] The matrix $G = \myE[g_{(\theta)}(\thetatrue)]$ exists, is
	finite, and has full rank.
	
	\item[C$_6$] The weighting matrix $M_T$ converges in probability to a
	constant matrix $M$ with full rank.
	
	\item[C$_7$] The matrix $\myE[\one{X \neq Y} \hspace{0.1mm} WW']$
	exists, is finite, and has full rank.
	
\end{enumerate}

As compared to the assumption of stationarity and ergodicity in B$_1$,
condition C$_1$ enforces stationarity and $\alpha$-mixing, which
implies ergodicity \citep[][Proposition 3.44]{white2014asymptotic}.
The stronger condition is essential, as asymptotic normality fails
generically under data generating processes with long memory
\citep{Beran1994}.  The moment constraints in C$_{2,Q}$ and C$_{2,E}$
can be weakened at the expense of stronger mixing conditions
\citep{hansen1992consistent}.  

\begin{theorem}[Asymptotic normality]  \label{thm:normality}
	For an optimal state-dependent quantile and a consistent GMM estimator
	of the form in \eqref{eq:GMM}, conditions A$_Q$, C$_1$, C$_{2,Q}$,
	C$_3$, C$_4$, C$_5$, and C$_6$ guarantee asymptotic normality with
	asymptotic covariance matrix 
	\begin{equation}  \label{eq:as.cov} 
		\left( G M^{-1} G' \right)^{-1} 
		\left( G M^{-1} \Sigma M^{-1} G' \right)
		\left( G M^{-1} G' \right)^{-1}.  
	\end{equation} 
	Analogously, for an optimal state-dependent expectile and a consistent
	GMM estimator conditions A$_E$, C$_1$, C$_{2,E}$, C$_3$, C$_4$, C$_5$,
	and C$_6$ yield the same conclusion.
\end{theorem} 

\begin{proof}
	The assumption of consistency along with $C_4$ guarantee that the GMM
	estimator $\thh$ in \eqref{eq:GMM} satisfies Definition 3.1 of
	\cite{Hansen1982}.  Therefore, it suffices to verify the conditions of
	Theorem 3.1 of \citet{Hansen1982}, namely, Assumptions 3.1--3.6, with
	conditions C$_1$, C$_3$, and C$_4$ covering Assumptions 3.1, 3.2, and
	3.3, respectively.  Conditions C$_{2,Q}$ or C$_{2,E}$ along with C$_4$
	and C$_5$ imply Assumption 3.4.  Lemma 3.2 in \cite{Hansen1982},
	C$_5$, and C$_6$ yield Assumption 3.6.
	
	We proceed to verify Assumption 3.5.  For ease of notation, let $g_t =
	g_t(\thetatrue)$ and let ${\cal I}_{t-j}$ be the $\sigma$-algebra
	generated by $g_{t-j}, g_{t-j-1}, \ldots$ \ Then $\myE[ g_t g_t']$
	exists and is finite by C$_{2,Q}$ or C$_{2,E}$ respectively.  The same
	conditions along with the mixing inequalities in Lemma 1.3 of
	\citet{ibragimov1962some} or Theorem 14.2 of
	\citet{davidson1994stochastic}, applied with $p = 2$ and $q = 4$,
	imply that $\myE[ g_t | {\cal I}_{t-j}] \to 0$ as $j \to \infty$ in
	mean square.  Letting $v_j = \myE[ g_t | {\cal I}_{t-j}] - \myE[ g_t |
	{\cal I}_{t-j-1}]$, it remains to be shown that $\sum_{j=0}^\infty
	\myE[v_j'v_j]^{1/2}$ is finite, which follows from the aforementioned
	mixing inequality in concert with the triangle and H\"older's
	inequalities.
	
	We may now invoke Theorem 3.1 of \citet{Hansen1982}, which shows that
	$\sqrt{T}(\thh - \thetatrue) \to \cN_p(0,V)$ as $T \to \infty$, where $V
	= (G M^{-1} G')^{-1} \, (G M^{-1} \Sigma M^{-1} G') \, (G M^{-1}
	G')^{-1}$, as stated.
\end{proof}

In particular, Theorems \ref{thm:consistency} and \ref{thm:normality}
guarantee the consistency and asymptotic normality of the first step
GMM estimator, for which the unit matrix serves as weighting matrix.
We proceed to apply the latter result in order to demonstrate the
asymptotic distribution \eqref{eq:asymnormal} of the efficient
two-step GMM estimator $\thh$ based on a consistent first step
estimator $\thh^1$ and an associated heteroskedasticity and
auto\-correlation consistent \citep[HAC,][]{Newey1987a} estimator
$\Sigma_T$ of the covariance matrix $\Sigma$, with $\Sigma_T^{-1}$
serving as weighting matrix.  This requires condition C$_7$, which in
the case of quantiles reduces to the standard assumption that the
matrix $\myE[WW']$ has full rank, given that A$_Q$ implies $X \neq Y$
almost surely.  In the case of expectiles C$_7$ generally is neither
necessary nor sufficient for $\myE[WW']$ to have full rank.

\begin{corollary}[Two-step GMM with HAC covariance estimator]  \label{cor:HAC}
	For an optimal \linebreak state-dependent quantile and a consistent
	GMM estimator\/ $\thh^1$, conditions A$_Q$, C$_1$, C$_{2,Q}$, C$_3$,
	C$_4$, and C$_5$ guarantee the consistency of the HAC estimator
	$\Sigma_T$ that is based on\/ $\thh^1$.  If furthermore condition
	C$_7$ holds true, the two-step GMM estimator $\thh$ is asymptotically
	normal with asymptotic covariance matrix $\left( G \Sigma^{-1} G'
	\right)^{-1}$.
	
	Analogously, for an optimal state-dependent expectile and a consistent
	GMM estimator\/ $\thh^1$, conditions A$_E$, C$_1$, C$_{2,E}$, C$_3$,
	C$_4$, and C$_5$ guarantee the consistency of the HAC estimator.
	$\rule{0mm}{3.25mm}$ If furthermore condition C$_7$ holds true, the
	two-step GMM estimator\/ $\thh$ is asymptotically normal with
	asymptotic covariance matrix $\left( G \Sigma^{-1} G' \right)^{-1}$.
\end{corollary}

\begin{proof}
	We first show that the HAC estimator $\Sigma_T$ is consistent for
	$\Sigma$ by verifying the conditions of Theorem 2 in
	\citet{Newey1987a}, namely, Assumptions (i), (ii), (iii), (iv), and
	(v).  Assumptions (i) and (ii) are guaranteed by C$_{2,Q}$ or
	C$_{2,E}$ along with C$_4$.  Assumption (iii) is immediate from C$_1$,
	and to verify (iv) we apply Theorem \ref{thm:normality}.  By our
	implementation choices, (v) is trivially satisfied.  Thus, the HAC
	estimator is consistent for $\Sigma$.  In the case of expectiles we
	have $\Sigma = \myE[ (\one{Y \le X} - m(S,\thetatrue))^2 \, (X-Y)^2 \,
	WW']$.  If $\delta' \Sigma \delta = 0$ for some $\delta \in
	\real^q$, then $(\one{Y \leq X} - m(S,\thetatrue))^2 (X-Y)^2
	|W'\delta|^2 = 0$ almost surely, which implies $\delta = 0$ in view of
	condition C$_7$, whence $\Sigma$ has full rank.  In the case of
	quantiles a similar argument applies.
	
	Therefore, $\Sigma_T^{-1}$ is consistent for $\Sigma^{-1}$, and we may
	apply Theorem \ref{thm:normality} to the efficient two-step estimator
	$\thh$ with weighting matrix $M_T = \Sigma_T^{-1}$.  Invoking
	\eqref{eq:as.cov} we see that $\thh$ is asymptotically normal with
	asymptotic covariance matrix $\left( G \Sigma^{-1} G' \right)^{-1}$.
\end{proof}

\footnotesize
\bibliographystyle{apalike}
\bibliography{Bib}

\end{document}